\documentstyle[floats,aps,pra,psfig,eqsecnum]{revtex}

\begin{document}
\title{Quantum interference in the fluorescence of a molecular system}
\author{Jin Wang${}^{1,2}$, H.M. Wiseman${}^{1,2}$, and Z.
Ficek${}^{1,3}$}
\address{${}^{1}$Centre for Laser Science and
Department of Physics, The University of Queensland, Brisbane,
Queensland 4072, Australia\\
${}^{2}$ School of Science, Griffith University, Nathan, Brisbane, Queensland
4111, Australia \\
${}^{3}$ Department of Applied Mathematics and Theoretical Physics,
The Queen's University of Belfast, Belfast BT7 1NN, Northern Ireland}
\date{\today}
\maketitle

\begin{abstract}
                 It has been observed experimentally [H.R. Xia, C.Y. Ye,
                 and S.Y. Zhu, Phys. Rev. Lett. {\bf 77}, 1032 (1996)]
                 that quantum interference between two molecular transitions
                 can lead to a suppression or enhancement of spontaneous emission.
                This is manifested in the fluorescent intensity as a function of 
                the detuning of the driving field from the two-photon resonance condition.
                 Here we present a theory which explains the observed variation
                 of the number of peaks with the mutual polarization of 
                 the molecular transition
                 dipole moments. Using master equation techniques we calculate
                 analytically as well as numerically the steady-state
                fluorescence, and find that the 
                number of peaks depends on the excitation process. If the molecule is
                 driven to the upper levels by a two-photon process, the fluorescent
                 intensity consists of two peaks regardless of the mutual polarization of 
                the transition dipole moments. If the excitation
                 process is composed of both a two-step one-photon process and a one-step,
                 two-photon process, then there are two peaks on
                 transitions with parallel dipole moments and three peaks on transitions with
                 antiparallel dipole moments. This latter case is in excellent agreement with
                 the experiment.

\end{abstract}


\newcommand{\beq}{\begin{equation}}
\newcommand{\eeq}{\end{equation}}
\newcommand{\bqa}{\begin{eqnarray}}
\newcommand{\eqa}{\end{eqnarray}}
\newcommand{\nn}{\nonumber}
\newcommand{\nl}[1]{\nn \\ && {#1}\,}
\newcommand{\erf}[1]{Eq.~(\ref{#1})}
\newcommand{\dg}{^\dagger}
\newcommand{\rt}[1]{\sqrt{#1}\,}
\newcommand{\smallfrac}[2]{\mbox{$\frac{#1}{#2}$}}
\newcommand{\half}{\smallfrac{1}{2}}
\newcommand{\bra}[1]{\langle{#1}|}
\newcommand{\ket}[1]{|{#1}\rangle}
\newcommand{\ip}[1]{\langle{#1}\rangle}
\newcommand{\sch}{Schr\"odinger }
\newcommand{\schs}{Schr\"odinger's }
\newcommand{\hei}{Heisenberg }
\newcommand{\heis}{Heisenberg's }
\newcommand{\bl}{{\bigl(}}
\newcommand{\br}{{\bigr)}}

\newcommand{\dbd}[1]{\frac{\partial}{\partial {#1}}}

\section{Introduction}

There have been a large number of theoretical studies on the effects of
quantum interference in atomic and molecular systems~\cite{ari}. This
phenomenon was
first suggested by Agarwal~\cite{aga} who showed that the spontaneous
emission from
a degenerate $V$-type three-level atom is sensitive to the mutual
orientation of the atomic dipole moments. If they are parallel a
suppression of spontaneous emission can appear and a part of the
population can be trapped in the excited levels. Similar predictions were
reported for other configurations of three- and multi-level atoms and
show that quantum interference can lead to many interesting effects such
as amplification without population inversion~\cite{har},
electromagnetically-induced transparency~\cite{bol}, phase dependent spectra
and population inversions~\cite{pat}, and ultranarrow spectral
lines~\cite{zhou}.

Zhu and Scully~\cite{zhu} and Lee {\it et al.}~\cite{lee} have shown that
in the case of a non-degenerate $V$-type atom
driven from an auxiliary level, quantum interference can lead to
the elimination of the central line in the fluorescence spectrum
when the driving field is tuned to the middle of the upper levels
splitting. This interesting effect suggests that quantum interference can
be used as a mechanism for controlling and even for suppression of
spontaneous emission.

In 1996, Xia {\it et al.}~\cite{xia} carried out the first experimental
investigation of constructive and destructive interference effects in
spontaneous emission. In the experiment they used sodium dimers, which can
be modeled as five-level molecular systems with a single ground level, two
intermediate and two upper levels, driven by a two-photon process from the
ground level to the upper doublet. By monitoring the fluorescence from the
upper levels they observed that the total fluorescent intensity, as a
function of two-photon detuning, is composed of two peaks on
transitions with parallel and three peaks on transitions with
antiparallel dipole moments. The observed variation of the number of peaks
with the mutual polarization of the dipole moments gives compelling
evidence for quantum interference in spontaneous emission.

It is our purpose in this paper to present a theoretical explanation
of the observed fluorescent intensity and, in particular, to explain
the variation of the number of the observed peaks with the mutual polarization
of the molecular dipole moments. We point out here that the previous
theoretical studies~\cite{zhu,lee} of quantum interference between two
transitions with parallel or antiparallel dipole moments have dealt with the
fluorescence {\em spectrum}. By contrast in the experiment, the {\em total} fluorescent
intensity,
as a function of two-photon detuning, was observed. Agarwal~\cite{gsa} has
provided an intuitive picture for the observed spontaneous emission
cancellation
in terms of interference pathways involving a two-photon absorption process.
Recently, Berman~\cite{ber} has shown that the experimentally observed
cancellation of spontaneous emission involving a two-photon absorption process
can be interpreted in terms of population trapping. Although a cancellation
of spontaneous emission is present with a two-photon excitation
process, no variation of the number of peaks with the polarization of
the dipole moments exist in the fluorescent intensity. In summary, no
explanation has been offered until now for the observed variation of the
number of peaks in the fluorescent intensity with the mutual polarization
of the transition dipole moments.

In this paper we consider a five-level system driven by a single-mode
coherent laser field, which models the experimental configuration set
up by Xia {\it et al.}~\cite{xia}. Working with the master equation of the
system, we calculate the steady-state fluorescent intensity as a function
of the
laser frequency for two different transitions from the upper levels to
intermediate levels.
One transition is in the visible region and has parallel dipole moments.
The other transition is in the ultraviolet and has antiparallel 
dipole moments. We assume that there is spontaneous emission from the upper
to the intermediate levels and thence to the ground level
so the dynamics of the system are restricted to these five levels.
In a real sodium molecule, the situation is more complex, with other
decay channels, and laser-field couplings between various real
states~\cite{wang}. However, we believe that our simple model does explain
the basic physical effects which have been observed in the experiment.

In Xia's paper~\cite{xia} the excitation of the upper states
is described as a two-photon process. As we will see later, the two
photon excitation process can only ever lead to two peaks in the
fluorescent intensity, independent of the mutual polarization of the
dipole moments. We show that the experimentally observed variation of
the number of peaks arises from the presence of an additional two-step,
one-photon excitation processes.

The paper is organized as follows. The master equation for the five-level
molecular system driven by a single mode laser field is derived and
analyzed in section II. The analytical and numerical results for the 
total fluorescent intensity for the two-photon coupling only are studied 
in section III. In section IV, we investigate the corresponding results when  
the system has both one- and two-photon coupling. We also examine the
approximations made and make comparisons with the experimental results.
A discussion is given in the concluding section V.

\section{The Master Equation}

\begin{figure}[bp]
        \center
\center

\centerline{\hbox{
\psfig{figure=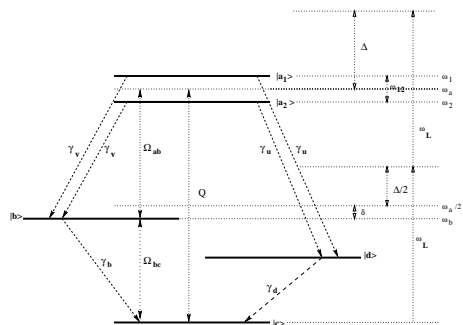,width=6cm}}}
        \caption{Energy level structure and couplings of the molecular system.}
        \label{fig:els}
\end{figure}

The energy-level scheme of the system we are considering is shown in
Fig.~\ref{fig:els}, in which we follow the notation of Ref.~\cite{xia}.
 The five-level molecule consists of two upper levels
$\left|a_{1}\right\rangle$ and $\left|a_{2}\right\rangle$,
two intermediate levels $\ket{b}$ and $\ket{d}$, and a single ground
level $\ket{c}$. The upper levels are separated by the frequency
$\omega_{12}$ which is much smaller than the frequencies $\omega_{1b}$ and
$\omega_{2b}$ of the $\ket{a_{1}}\rightarrow \ket{b}$ and
$\ket{a_{2}}\rightarrow \ket{b}$ transitions and the frequencies
$\omega_{1d}$ and $\omega_{2d}$ of the $\ket{a_{1}}\rightarrow \ket{d}$ and
$\ket{a_{2}}\rightarrow \ket{d}$ transitions. As in the
sodium dimers used in the Xia's experiment~\cite{xia}, we assume that the
frequencies $\omega_{1b}$ and $\omega_{2b}$ are significantly different
from the frequencies $\omega_{1d}$ and $\omega_{2d}$. The transitions
$\ket{a_{1}},\ket{a_{2}}\rightarrow \ket{b}$ correspond to the visible
region, whereas the transitions $\ket{a_{1}},\ket{a_{2}}\rightarrow\ket{d}$
correspond to the uv region.

In the molecule, the one-photon transitions
$\ket{a_{1}},\ket{a_{2}}\rightarrow\ket{b},\ket{d}\rightarrow \ket{c}$
are connected by electric dipole moments,
whereas the transition $\ket{a_{1}}\rightarrow\ket{a_{2}}$ and
the two-photon transitions $\ket{a_{1}}, \ket{a_{2}}\rightarrow \ket{c}$
are forbidden in the electric dipole approximation. The molecular
dipole moments can have different orientations (polarizations) and
two dipole moments which are close in frequency can interfere with each other
if they are not orthogonal. In the experiment, a destructive interference
was observed between two transitions, $\ket{a_{1}}\rightarrow \ket{b}$ and
$\ket{a_{2}}\rightarrow \ket{b}$, with parallel dipole moments,
and a constructive interference was observed between transitions
$\ket{a_{1}}\rightarrow \ket{d}$ and $\ket{a_{2}}\rightarrow \ket{d}$
with antiparallel dipole moments. 

In order to quantify the mutual
orientations of the transition dipole moments, we introduce a parameter
\bqa
p &=& \frac{\vec\mu_{ij}\cdot
\vec\mu_{kl}}{|\vec\mu_{ij}||\vec\mu_{kl}|}, \qquad {ij} \neq {kl} ,
\eqa
where $\vec\mu_{ij}$ is the matrix element of the transition dipole
moment between $\ket{i}$ and $\ket{j}$ levels. Using the subscripts
$u$ and $v$ to denote the ultraviolet and visible transitions in the
experiment,
we have $p_u=1$ (parallel dipole moments), while $p_v=-1$ (antiparallel 
dipole moments). 

For simplicity we will assume that the magnitude of the interfering dipole 
moments are the same. Thus, the upper doublet decays to level $\ket{b}$
at rate $\gamma_v=\gamma_{1b}=\gamma_{2b}$ and to level 
$\ket{d}$ at rate $\gamma_u=\gamma_{1d}=\gamma_{2d}$.
Here 
$u$ and $v$ again refer to visible and ultraviolet. 
The intermediate levels $\ket{b}$ and $\ket{d}$ decay to the ground level
$\ket{c}$
at rates $\gamma_b$ and $\gamma_d$ respectively.

The system is driven by a single-mode tunable laser of frequency
$\omega_L$. In the experiment the dye laser was coupled to the two-photon
transition $\ket{c}\rightarrow \ket{a_{1}}, \ket{a_{2}}$ in order to
avoid the Doppler effect (which we ignore in our analysis). 
Here, we must ask a question whether the
two-photon coupling in the experiment was the only coupling of the
laser to the system. It is stated in the experimental paper~\cite{xia}
that the two-photon transition in sodium dimers was enhanced by a
near-resonant intermediate level, indicating that the laser could also couple 
the ground state $\ket{c}$ to the upper states $\ket{a_1},\ket{a_2}$ via 
cascaded one-photon transitions. Here, to avoid introducing an extra level,
we take the 
near-resonant intermediate level to be $\ket{b}$, so the laser can also 
produce a two-step one-photon transition
$\ket{c}\rightarrow \ket{b}$ then
$\ket{b}\rightarrow \ket{a_{1}},\ket{a_{2}}$. In our opinion this
channel of the excitation was possible in the experiment as the
one-photon transitions in the molecule are in the visible region and
their dipole moments  are parallel \cite{xia}. We will see later that the
presence of this channel of excitation will be crucial in the
explanation of the experimentally observed fluorescent intensity
profile. With only two-photon excitation   quantum interference
can be observed but the fluorescent intensity exhibits two peaks 
(as a function of laser detuning)
regardless of the mutual orientation of the transition dipole moments.
The three peak structure of the fluorescent intensity observed in
the experiment in the uv region can only result from the presence of the
two-step one-photon channel.

We calculate the steady-state intensity of the fluorescence
from the upper doublet to the intermediate levels as follows. The intensity is
proportional to the normally ordered first-order correlation function
of the scattered field
\bqa
I\left(\vec{r},t\right) &\propto& \left\langle
\vec{E}^{\left(-\right)}\left(\vec{r},t\right)\cdot
\vec{E}^{\left(+\right)}\left(\vec{r},t\right)\right\rangle ,
\eqa
where $\vec{E}^{\left(+\right)}\left(\vec{r},t\right)$ is the
positive frequency part of the electric field operator at a point
$\vec{r}$ in the far-field zone of the system outside the driving
laser field. In terms of the density matrix elements of the system the
scaled steady-state $(t\to\infty)$ intensity on the ultraviolet and visible
transitions
is 
\beq \label{int}
I_{u/v} =\gamma_{u/v}\left(
\rho_{11}+\rho_{22}+2p_{u/v}{\rm Re}\rho_{12}\right) .
\eeq
Here $\rho_{11}$ and  $\rho_{22}$ are the
steady-state populations of the level $\ket{a_{1}}$ and $\ket{a_{2}}$, and
$\rho_{12}$ is the stead-state coherence between them.

We find steady-state values of the populations and coherences from the
master equation of the system. The master equation can be written in
the Lindblad form~\cite{Lin76} as
\beq \label{me1}
\dot{\rho} = {\cal L}_{\rm rev}\rho + {\cal L}_{\rm
irr}\rho ,
\eeq
where the reversible and irreversible terms are, respectively
\bqa \label{Ham}
{\cal L}_{\rm rev}\rho = -i[H,\rho],
\eqa
and
\bqa
{\cal L}_{\rm irr}&=& \gamma_v\left(1+p_v\right){\cal
D}\left[\ket{b}(\bra{a_1}+\bra{a_2})/\sqrt{2}\right]
+\gamma_v \left(1-p_v\right){\cal
D}\left[\ket{b}(\bra{a_1}-\bra{a_2})/\sqrt{2}\right]\nn\\
&&+\,\gamma_u\left(1+p_u\right){\cal
D}\left[[\ket{d}(\bra{a_1}+\bra{a_2})/\sqrt{2}\right]
+\gamma_u\left(1-p_u\right){\cal
D}\left[\ket{d}(\bra{a_1}-\bra{a_2})/\sqrt{2}\right]\nn\\ &&+\,
\gamma_b{\cal D}[\ket{c}\bra{b}]+\gamma_d{\cal D}[\ket{c}\bra{d}]   \\
&=& \gamma_v{\cal D}\left[\ket{b}(\bra{a_1}-\bra{a_2})\right] 
+ \gamma_u{\cal D}\left[\ket{b}(\bra{a_1}+\bra{a_2})\right] 
+\gamma_b{\cal D}[\ket{c}\bra{b}]+\gamma_d{\cal D}[\ket{c}\bra{d}].
\eqa
Here, ${\cal D}$ is a superoperator defined for arbitrary operators
$A$ and $B$ as
\beq
{\cal D}[A]B \equiv ABA\dg - \half \{A\dg A,B\}.
\eeq
Taking the ground state to have zero energy, the Hamiltonian operator in
\erf{Ham} 
(working in units where $\hbar=1$) can be split as $H=H_{0}+H_1$, where
\bqa \label{H0}
H_0&=&2\omega_L\ket{a_1}\bra{a_1}+
2\omega_L\ket{a_2}\bra{a_2}+\omega_L\ket{b}\bra{b}+\omega_d\ket{d}\bra{d} 
\eqa
is approximately equal to the Hamiltonian of the molecular system, and 
\bqa \label{inter}
H_1&=&\left[ \Omega_{bc}\ket{b}\bra{c}e^{-i\omega_{L}t}+{\rm 
H.c.}\right]+\left[\Omega_{ab}\left(\ket{a_1}+\ket{a_2}\right)
\bra{b}e^{-i\omega_{L}t}+{\rm H.c.}\right]\nn\\ 
&& + \left[Q\left(\ket{a_1}+\ket{a_2}\right)
\bra{c}e^{-i2\omega_{L}t}+{\rm H.c.}\right]\nn\\ 
&& +(\omega_{1}-2\omega_L)\ket{a_1}\bra{a_1}
+(\omega_{2}-2\omega_L)\ket{a_2}\bra{a_2}+(\omega_{b}-\omega_L)\ket{b}\bra{b}
+ (\omega_d - \omega_L)\ket{d}\bra{d}
\eqa
includes the interaction with the laser field plus corrections to 
$H_0$ to reproduce the full molecular Hamiltonian.

  The first and second terms in
Eq.~(\ref{inter}) describe the interaction of the classical laser field
with electric dipole moments of the one-photon transitions
$\ket{c}\rightarrow \ket{b}$ and $\ket{b}\rightarrow
\ket{a_{1}},\ket{a_{2}}$, respectively. The strengths with which these
transitions are driven are characterized by the one-photon Rabi
frequencies $
\Omega_{bc} = \frac{1}{2}\vec{\mu}_{bc}\cdot \vec{E}_{L}$, 
and
$\Omega_{ab} = \frac{1}{2}\vec{\mu}_{ba_{1}}\cdot \vec{E}_{L}
=\frac{1}{2}\vec{\mu}_{ba_{2}}\cdot \vec{E}_{L}$,
where $\vec{E}_{L}$ is the amplitude of the laser field.

The third term in Eq.~(\ref{inter}) describes the two-photon coupling of
the laser field to the system with the two-photon Rabi frequency
\bqa
Q = \sum_{m} \frac{1}{2}\frac{\mu_{mc}\mu_{ma_{1}}E_{L}^{2}}
{\omega_{L}-\omega_{mc}}
=\sum_{m} \frac{1}{2}\frac{\mu_{mc}\mu_{ma_{2}}E_{L}^{2}}
{\omega_{L}-\omega_{mc}},
\eqa
where $E_{L}=|\vec{E}_{L}|$. This is due to transitions via the
intermediate virtual levels labelled $m$ here.

Because of the external driving the 
elements of the system state matrix $\rho$ satisfy equations of 
motion 
containing explicit time-dependent factors of the complex exponential 
type.  
These can be removed by moving to the interaction picture with 
respect to $H_{0}$. The remaining Hamiltonian $H_1$ becomes
\bqa
H_I(t)&=&({\omega_{12}}/{2}-\Delta)\ket{a_1}\bra{a_1}\nn\\ 
&&+(-{\omega_{12}}/{2}-\Delta)\ket{a_2}\bra{a_2}\nn\\ 
&&+(-\Delta/2-\delta)\ket{b}\bra{b}\nn\\ 
&&+\left[\Omega_{ab}\left(\ket{a_1}+\ket{a_2}\right)\bra{b}+{\rm
H.c.}\right]\nn\\ 
&&+\left[Q\left(\ket{a_1}+\ket{a_2}\right)\bra{c}+{\rm H.c.}\right]\nn\\ 
&&+\left[\Omega_{bc}\ket{b}\bra{c}+{\rm H.c.} \right].\label{Hint}
\eqa 
Although this is written as $H_{I}(t)$ it is actually 
time-independent because of the judicious choice of $H_{0}$. 
Here $\Delta = 2\omega_L - \omega_a$ is the detuning between the two-photon
laser 
frequency $2\omega_L$ and the mean frequency of the upper levels relative
to the ground level $\omega_a = (\omega_1+\omega_2)/2$. The one-photon
detuning 
$\delta = \omega_L - \Delta/2 -\omega_b= \omega_a/2-\omega_b$ is the gap
between the energy of 
level $\ket{b}$ and the half way position from the ground level $\ket{c}$
to the mean of the upper levels $\ket{a_1}$ and $\ket{a_2}$. Moving to the
interaction picture does not affect the irreversible 
terms so the new master equation is 
\beq \label{me2}
\dot\rho = {\cal L}_{\rm irr}\rho - i[H_{I},\rho].
\eeq
The stationary solution satisfying $\dot{\rho}=0$ can be found numerically
and, 
in certain limits, analytically. We consider separately the case of
two-photon 
coupling only, and one- and two-photon coupling.

\section{Two-Photon Coupling Only}

The case where the upper pair of levels is excited only by two-photon
transitions via virtual intermediate levels is found by setting $\Omega_{ab}$
and $\Omega_{bc}$ in the interaction Hamilltonian (\ref{Hint}) equal to zero.
The two-photon driving parametrized by $Q$ is the only sort of driving
mentioned
in the experimental paper~\cite{xia}.

\subsubsection{Analytical Solution}
\label{sec:2phonly}

We first consider an analytical solution. This is possible in the
weak-field limit 
where $Q$ is much smaller than the decay rates in the system. For the 
experimentally relevant mutual polarizations $p_v = 1$, $p_u = -1$, the 
equations of motion are greatly 
simplified if we make the assumption that $\gamma_u = \gamma_v$. 
That is, we assume that the decay rates of the upper levels on the 
ultraviolet and visible transitions are equal. We therefore define a new 
parameter $\gamma_a = \gamma_u = \gamma_v$. 

\begin{figure}[tbp]
        \center
\centerline{\hbox{
\psfig{figure=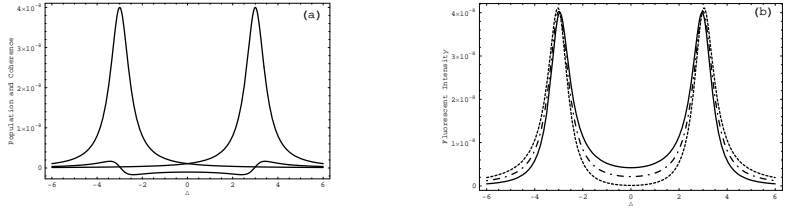,width=12cm}}}
        \caption{Analytical results for 2-photon coupling only. 
(a) shows populations ($\rho_{11}$ peaked to the left, $\rho_{22}$ peaked
to the right) 
and coherence (${\Re}\rho_{12}$, below the axis)  of the excited states.
(b) shows the total fluorescent intensity in units of the spontaneous
emission rate. 
The solid line shows the intensity on the ultraviolet transition ($p_u=-1$), 
the dashed line
the intensity on the visible transition ($p_v=1$), and 
the dash-dot line the hypothetical intensity for a transition with
orthogonal dipole moments 
($p=0$).
The parameters are $\Omega_{ab}=\Omega_{bc}=0,
Q=10^{-4},\omega_{12}=6,\delta=0,\gamma_u=\gamma_v=0.5,
\gamma_b=1$. The two-photon detuning $\Delta$ is plotted in units 
of $\gamma_u+\gamma_v$. }
        \label{fig:2ph}
\end{figure}

Under these assumptions, it is easy to show that the master equation 
(\ref{me2}) leads to the following steady-state values
of the upper level populations and coherences 
\bqa
\rho_{11}&=&\frac{Q^2}
{(\Delta+{\omega_{12}}/{2})^2+\gamma_a^2} ,\\
\rho_{22}&=&\frac{Q^2}
{(\Delta-{\omega_{12}}/{2})^2+\gamma_a^2} ,\\
{\rm Re}\rho_{12}&=&\frac{Q^2[\Delta^2-(\omega_{12}/2)^2+\gamma_a^2]}
{[(\Delta+{\omega_{12}}/{2})^2+\gamma_a^2][(\Delta-{\omega_{12}}/{2})^2+\gamma_a^2]}.
\eqa
These are shown in Fig.~\ref{fig:2ph}(a) as a function of $\Delta$.

This analysis predicts that the populations
and coherence exhibit peaks at $\Delta =\pm
\omega_{12}/2$, corresponding to the two-photon resonances
of the laser field with the $\ket{c}\rightarrow\ket{a_{1}}$ and
$\ket{c}\rightarrow\ket{a_{2}}$ transitions. In Fig.~\ref{fig:2ph}(b),
we
plot the
fluorescent intensity as a function of $\Delta$ for the $p_{v}=1$
and $p_{u}=-1$ transitions. It is seen that there are two
peaks located at $\Delta =\pm \frac{1}{2}\omega_{12}$, the amplitudes of
which are not sensitive to $p$. The intensity is sensitive to
$p$ only about $\Delta =0$ and can be almost completely suppressed for
$p_{v}=1$ transitions. This confirms the earlier prediction by
Agarwal~\cite{gsa}
that the two-photon excitation process involving the $\ket{a_{1}}$
and $\ket{a_{2}}$ levels can lead to cancellation of spontaneous
emission to the level $\ket{b}$. The cancellation of the
fluorescence at $\Delta =0$ also confirms the prediction by
Berman~\cite{ber}
that the suppression of the fluorescence can be explained in terms of
dark states and coherent population trapping.

For $p_{v}=1$ the
fluorescent intensity~(\ref{int}) can be written as
\beq
I_{v} = 2\gamma_{v}\rho_{ss} ,
\eeq
where $\rho_{ss}=\bra{s}\rho \ket{s}$ is the population of the symmetric
$\ket{s} = \left(\ket{a_{1}}+\ket{a_{2}}\right)/\sqrt{2}$ combination of
the upper levels. The suppression of the fluorescence at $\Delta =0$ indicates that the
state $\ket{s}$ is almost unpopulated in the steady-state. This implies
that the population is trapped between other molecular levels, including
the antisymmetric state $\ket{a}=\left(\ket{a_{1}}-\ket{a_{2}}\right)/\sqrt{2}$, with the
state $\ket{s}$ being a dark state of the system.

\subsection{Numerical Results}

As noted before, in the experiment~\cite{xia} three peaks were observed on
the transitions
with antiparallel dipole moments. However, as it is seen from
Fig.~\ref{fig:2ph}(b),
the weak-field theory does not predict three peaks for the $p_{u}=-1$
transitions. The reason is that the magnitude of the coherence term 
$\rho_{12}$ is small compared to the magnitude of the population terms 
so that it is unable to build up a third peak in the
middle. The coherence term is necessarily small because there is no detuning 
$\Delta$ at which both populations are large, and the coherence term is 
limited in magnitude by
\beq \label{ineq}
|\rho_{12}|^2 \leq \rho_{11}\rho_{22}.
\eeq

To prove that the lack of a third peak is not a result of the assumptions 
made in deriving the analytical results we have also studied numerically 
the steady state of the master equation (\ref{me2}) with the one-photon 
Rabi frequencies set to zero. This can be done by calculating the equations 
of motion for the density matrix elements and using matrix inversion
techniques. 
It can be done more easily using the
direct symbolic representation of the master equation (\ref{me2}) 
which is possible in the quantum optics toolbox for
matlab~\cite{szeta}. We find that, even in the strong field limit, 
and even with $\gamma_u \neq \gamma_v$, it is not possible to 
produce a third peak in the fluorescence profile.

From these analytical and numerical results we conclude that 
as well as the two-photon excitation process there must be
some other processes involved in the dynamics of the system.
The obvious candidate is a two-step one-photon process.

\section{One- and Two-Photon Coupling}

To include one-photon coupling we now consider the case where $\Omega_{bc}$ 
and $\Omega_{ab}$ are nonzero. To include two-photon coupling we actually do
 not need to have $Q$ nonzero. That is because, as we will show, there is a 
 regime in which level $\ket{b}$ acts as a virtual level with almost no
real population. In this limit, the two step one-photon process becomes
equivalent to a two-photon process. The relative strength of the two- and
one-photon couplings is given by a parameter $\alpha = \gamma_b/\gamma_a$,
to be discussed later. Thus, for simplicity, we set $Q = 0$.

\subsection{Analytical Results}

To obtain analytical results we must consider the equations of motion for the
density matrix elements. The master equation~(\ref{me2}), in general,
leads to a system of twenty five equations of motion for the density
matrix elements. Because of the assumption of large non-degeneracy
between the intermediate levels $\ket{b}$ and $\ket{d}$, the
coherences $\rho_{cd}, \rho_{da_{1}}, \rho_{bd}$ and $\rho_{da_{2}}$ are
not coupled to the driving field, and then the system of equations splits
into two subsystems: one of seventeen equations of motion directly
coupled to the driving field and the other of eight equations of
motion not coupled to the driving field. It is not difficult to show
that the steady-state solutions for the eight density matrix elements
are zero and therefore we limit our considerations to the seventeen
equations which, after applying the trace property (Tr$\rho =1$),
reduce to a system of sixteen coupled linear inhomogeneous equations.

As in the case of Sec.~\ref{sec:2phonly}, for the physical parameters
$p_v=1$, $p_u=-1$, the equations are simplified if $\gamma_u=\gamma_v$. 
Under this assumption, and substituting $\gamma_a$ for  both $\gamma_u$ 
and $\gamma_v$, the relevant density matrix elements obey the following 
coupled equations
\bqa
\dot{\rho}_{11}&=&
-2\gamma_a\rho_{11}-i\Omega_{ab}\left(\rho_{b1}-\rho_{1b}\right) ,\\
\dot{\rho}_{22}&=&
-2\gamma_a\rho_{22}-i\Omega_{ab}\left(\rho_{b2}-\rho_{2b}\right) ,\\
\dot{\rho}_{bb}&=&
-\gamma_b\rho_{bb}+\gamma_a\left(\rho_{11}+\rho_{22}
+\rho_{12}+\rho_{21}\right) \nn \\
&-&i\Omega_{bc}\left({\rho}_{cb}-{\rho}_{bc}\right)
+i\Omega_{ab}\left({\rho}_{b1}-{\rho}_{1b}\right)
+i\Omega_{ab}\left({\rho}_{b2}-{\rho}_{2b}\right) ,\\
\dot{\rho}_{cc}&=&
\gamma_b\rho_{bb}+\gamma_d(1-\rho_{11}-\rho_{22}-\rho_{bb}-\rho_{cc})
+i\Omega_{bc}\left(\rho_{cb}-\rho_{bc}\right),\\
\dot{\rho}_{12}&=&
-\left(2\gamma_a+i\omega_{12}\right)\rho_{12}
-i\Omega_{ab}\rho_{b2}+i\Omega_{ab}\rho_{1b},\\
\dot{\rho}_{b2}&=&
-\left[\left(\gamma_a+\gamma_b/2\right)
-i\left(\delta-\Delta/2-\omega_{12}/2\right)\right]\rho_{b2}
-i\Omega_{bc}\rho_{c2}-i\Omega_{ab}\rho_{12}+i\Omega_{ab}\rho_{bb}-i\Omega_{ab}\rho_{22} ,\\
\dot{\rho}_{b1}&=&
-\left[\left(\gamma_a+\gamma_b/2 \right)
-i\left(\delta-\Delta/2+\omega_{12}/2\right)\right]\rho_{b1}
-i\Omega_{bc}\rho_{c1}-i\Omega_{ab}\rho_{12}+i\Omega_{ab}\rho_{bb}-i\Omega_{ab}\rho_{11},\\
\dot{{\rho}}_{bc}&=&
-\left[\gamma_b/2 -i(\delta+\Delta/2)\right]{\rho}_{bc}-i\Omega_{bc}
\left(\rho_{cc}-\rho_{bb}\right)
-i\Omega_{ab}\left({\rho}_{1c}+{\rho}_{2c}\right), \\
\dot{\rho}_{1c}&=& \left[-i(\omega_{12}/2-\Delta)-\gamma_a\right]\rho_{1c}
                                  -i\Omega_{ab}\rho_{bc}-\gamma_a\rho_{1c}+i\Omega_{bc}\rho_{1b},\\
\dot{\rho}_{2c}&=& \left[i(\omega_{12}/2+\Delta)-\gamma_a\right]\rho_{2c}
                                        +i\Omega_{ab}\rho_{bc} -\gamma_a
                                         \rho_{2c}-i\Omega_{bc}\rho_{1b}.
\eqa
To proceed further we make the weak-field assumption that $\Omega_{ab}$ 
and $\Omega_{bc}$ are small compared with the decay rates. We will show 
later that the same qualitative results can be obtained when this assumption, 
and the assumption $\gamma_u=\gamma_v$, are relaxed.

\begin{figure}[bp]
        \center
\center
\centerline{\hbox{
\psfig{figure=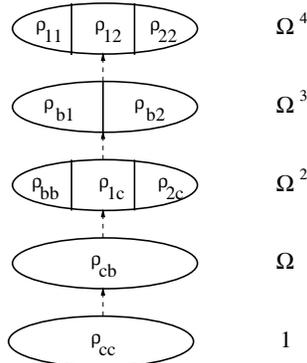,width=4cm}}}
        \caption{Diagram showing the method for solving 
        the steady state Master equation  under the weak field assumption. 
The symbols on the right represent the order of the matrix elements.}
        \label{fig:me}
\end{figure}

Under the weak field assumption we can order the matrix elements by how
they scale with 
$\Omega \sim \Omega_{bc},\Omega_{ab}$ as shown in Fig.~\ref{fig:me}. 
The simplifications result 
from keeping only the lowest order terms in the above equations of motion 
for the state matrix elements. The steady-state solutions can then be
obtained 
by setting the time derivatives to zero and solving the equations 
in the order as shown in Fig.~\ref{fig:me}.
We find that the upper level populations are given by 
\bqa
\rho_{11}&=&\frac{\Omega_{ab}^2\Omega_{bc}^2}
{[(\Delta/2+\delta)^2+{\gamma_b^2}/{4}]
{[(\Delta-{\omega_{12}}/{2})^2+\gamma_a^2]}} \label{pop1} ,\\
\rho_{22}&=&\frac{\Omega_{ab}^2\Omega_{bc}^2}
{[(\Delta/2+\delta)^2+{\gamma_b^2}/{4}]
{[(\Delta+{\omega_{12}}/{2})^2+\gamma_a^2]}} \label{pop2} .
\eqa

\begin{figure}[tbp]
        \center
\center
\centerline{\hbox{
\psfig{figure=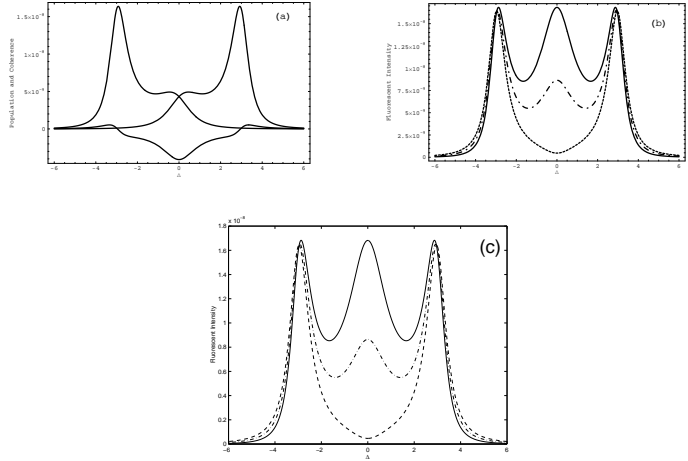,width=10cm}}}
        \caption{Weak field results for one- and two-photon driving. 
(a) shows analytical populations and coherences of the the excited states,
as in 
Fig.~\ref{fig:2ph}(a).
(b) shows the analytical results and (c) the numerical results for
the total fluorescent intensity in units of the spontaneous emission rate. 
The different line styles are as in Fig.~\ref{fig:2ph}(b).
        The parameters are
        $\Omega_{ab}=\Omega_{bc}=0.01,Q=0,\omega_{12}=6,\delta=0,
\gamma_u=\gamma_v=0.5,\gamma_b=1$.
 The two-photon detuning $\Delta$ is plotted in units of $\gamma_u+\gamma_v$.}
        \label{fig:1ph}
\end{figure}

This result predicts that, for large enough level splitting $\omega_{12}$, 
the population of both of the upper pair states 
have two distinct peaks as a function of laser detuning $\Delta$. 
This is illustrated in Fig.~\ref{fig:1ph}(a). The 
first peak is centered at $\Delta=\pm\omega_{12}/2$ for 
$\rho_{11}$ or $\rho_{22}$ respectively.  At this detuning the 
two-photon transition from $\ket{c}$ to $\ket{a_{1}}$ or $\ket{a_{2}}$ is
resonant, 
explaining the peak. The second peak is at $\Delta=-2\delta$. This is 
the resonance condition for the transition from $\ket{c}$ to $\ket{b}$, 
as seen in Fig.~\ref{fig:els}. This central 
peak results from two stepwise one-photon transitions, the first populating 
level $\ket{b}$ and the second exciting from $\ket{b}$ to $\ket{a}$. The
populating of 
level $\ket{b}$ at this laser frequency is evident from the steady-state 
result
\beq
\rho_{bb}=\frac{\Omega_{ab}\Omega_{bc}}
{[(\Delta/2+\delta)^2+{\gamma_b^2}/{4}]} .\\
\eeq

The upper-states 
coherence is considerably more complicated, and is given in full in the
Appendix. From the denominators in the expression given there, 
it is evident that 
$\rho_{12}$ may have many peaks and
 this is also illustrated in Fig.~\ref{fig:1ph}(a). To discover the
physical meaning out of such a 
complicated expression, we consider the limit of large 
splitting where $\omega_{12}$ is much larger than all other rates or 
frequencies. We then consider the behavior of  $\rho_{12}$ at 
the positions of its peaks, and keep only the leading contributions 
there. It turns out that only one peak survives this simplification:
\beq 
{\rm Re}\rho_{12}\simeq \frac{-\Omega_{ab}^2\Omega_{bc}^2}
{(\omega_{12}/2)^2\left[(\Delta/2+\delta)^2+
(\gamma_b/2)^{2}\right]}.\label{r1}
\eeq

From this expression it is evident that the coherence $\rho_{12}$ also
exhibits the
resonance at $\Delta = -2\delta $, and its magnitude is comparable to the 
magnitude of population terms. This is possible because the populations
$\rho_{11}$ 
and $\rho_{22}$ both have peaks at $\Delta = -2\delta$, resulting from
two-step 
one-photon transitions. Thus the inequality in \erf{ineq} allows the
coherence 
(\ref{r1}) to have a peak here also, unlike the case with only two-photon
transitions.

Assuming again that $\omega_{12} \gg \gamma_a, \gamma_b, \delta$, 
the peaks in the populations (\ref{pop1}) and (\ref{pop2}) are 
well-separated Lorentzians. Then using Eq.~(\ref{r1}), the 
fluorescent intensity for the ultraviolet and visible transitions can be
approximated as
\beq \label{int2}
I_{u/v} =
\frac{16\gamma_{u/v}\Omega_{ab}^2\Omega_{bc}^2}{\omega_{12}^2}
\left[\frac{1}{(\Delta-\omega_{12}/2)^2+\gamma_a^2}+
\frac{1}{2}\frac{\left(1-p_{u/v}\right)}{(\delta+\Delta/2)^2+
{(\gamma_b/2)}^2}+
\frac{1}{(\Delta+\omega_{12}/2)^2+\gamma_a^2}\right].
\eeq
In this limit the fluorescent intensity contains three Lorentzians located
at $\Delta =\pm \omega_{12}/2$ and $\Delta = -2\delta$.  
This is seen in the complete analytical solution for the fluorescence, 
from Eqs.~(\ref{pop1}), (\ref{pop2}), and (\ref{pop12}), plotted in
Fig.~\ref{fig:1ph}(b). 
The amplitude of
the peak at $\Delta =-2\delta$ strongly depends on the mutual
polarization of the dipole moments. The peak is absent in the
intensity $I_{v}$ observed in the visible region with $p_v=1$.
For the fluorescent intensity $I_{u}$ observed in the uv region with
$p_u=-1$, the amplitude of the peak is enhanced. The strong dependence
of the amplitude of the central peak on the mutual orientation of the
molecular dipole moments is precisely the effect observed in the experiment.
We emphasize again that the presence of the central peak in the
fluorescent intensity results from the coupling of the driving laser
to the one-photon transitions.

\subsection{Numerical results}

Having illustrated the role of the one- and two-photon excitations
in the weak-field limit, we now find the fluorescent intensity without
making any simplifying assumptions in our model. 
In this case it is not possible to obtain analytical
solutions and therefore we use numerical methods to find
stationary values of the density matrix elements of the system.
Once again, this is easy using the symbolic representational power 
of the quantum optics toolbox for matlab \cite{szeta}.
We first verify the correctness of the numerical technique 
by reproducing the weak-field analytical results. 
This is shown in Fig.~\ref{fig:1ph}(c).

\begin{figure}[tbp]
        \center
\center
\centerline{\hbox{
\psfig{figure=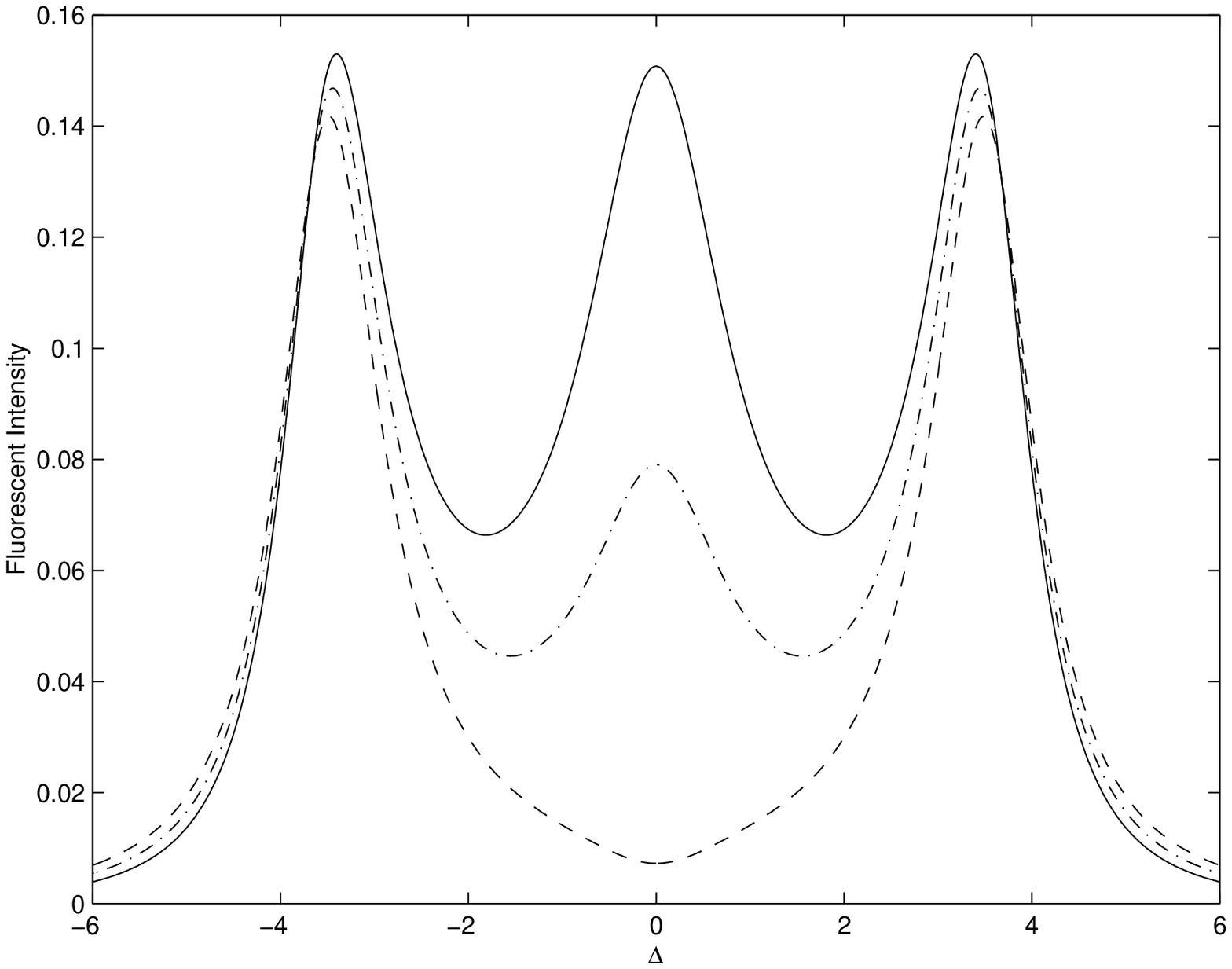,width=6cm}}}
        \caption{Total fluorescent intensity in units of the spontaneous emission
rate
for strong one- and two-photon coupling. The different line styles are as
in Fig.~\ref{fig:2ph}(b).
The parameters are
$\Omega_{ab}=\Omega_{bc}=1,Q=0,\omega_{12}=6,\delta=0,
\gamma_u=\gamma_v=0.5,\gamma_b=0.15$.
 The two-photon detuning $\Delta$ is plotted in units of
$\gamma_u+\gamma_v$.}   
        \label{fig:strong}
\end{figure}

In Fig.~\ref{fig:strong}, we plot the fluorescent intensity for a strong
driving field. It is seen that the fluorescent intensity exhibits
the same behavior as that for the weak driving field, shown in
Fig.~\ref{fig:1ph}, despite the fact that the solutions have been
derived in different regimes. 

\begin{figure}[tbp]
        \center
\center

\centerline{\hbox{
\psfig{figure=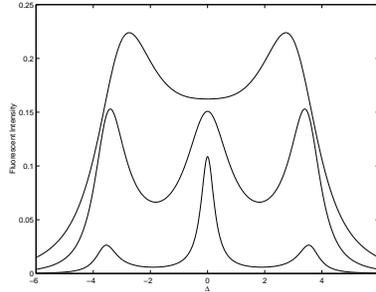,width=5cm}}}
\center

        \caption{Total fluorescent intensity in units of the spontaneous emission
rate
for strong one- and two-photon coupling. Only the ultraviolet ($p_u=-1$)
transition 
is plotted, but $\gamma_b = \alpha \gamma_u$ is varied. The values of
$\alpha$ for the three
curves are, from top to bottom, $2$, $0.3$, and $0.02$. The other
parameters are
$\Omega_{ab}=\Omega_{bc}=1,Q=0,\omega_{12}=6,\delta=0,
\gamma_u=\gamma_v=0.5$.
 The two-photon detuning $\Delta$ is plotted in units of $\gamma_u+\gamma_v$.}
        \label{fig:alpha}
\end{figure}

 The experimentally observed fluorescent intensity was asymmetric
about $\Delta =0$. There are few factors which could contribute towards
the observed asymmetry. For example, the decay rates from
the two upper levels to the intermediate levels could be unequal. 
A simpler reason could be that the central peak is not exactly at
$\Delta =0$. The analytical solution (\ref{int}) predicts the
central peak to be at $\Delta = -2\delta$ and the condition of $\delta = 0$
implies that the energy of the level $\ket{b}$ is exactly half of the
mean energy of the upper levels.
There is no reason to expect this condition to be satisfied in the real
molecule, and in fact it appears from the experimental results that
$\delta$ is positive. Fig.~\ref{fig:alpha} shows the effect
of a non-zero $\delta$ on the fluorescence profile for a strong driving
field. 

\begin{figure}[tbp]
\center
\centerline{\hbox{
\psfig{figure=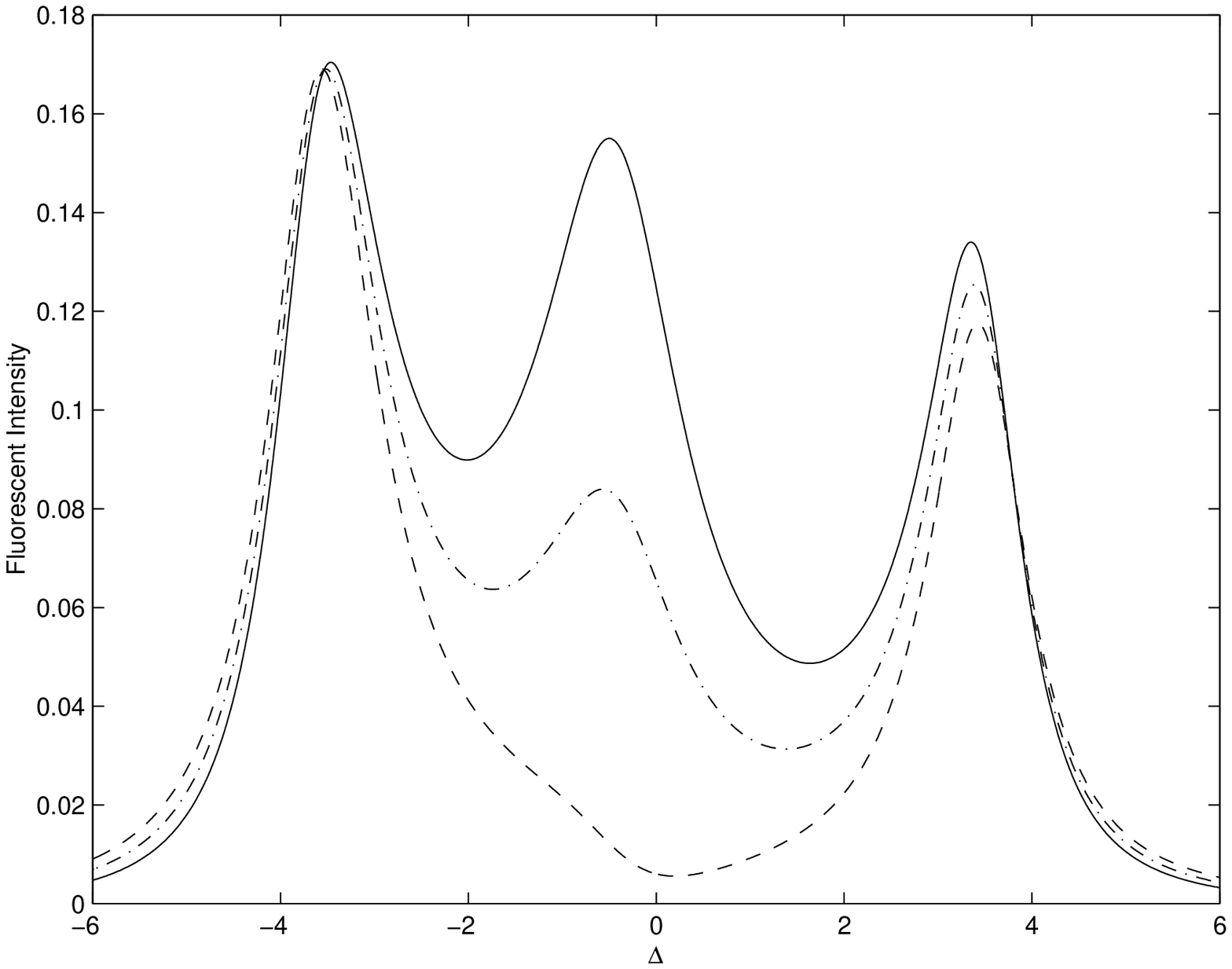,width=6cm}}}
\center 
        \caption{Total fluorescent intensity in units of the spontaneous emission
rate
for strong one- and two-photon coupling, with non-zero 
detuning $\delta$ of the intermediate level $\ket{b}$.
 The different line styles are as in Fig.~\ref{fig:2ph}(b).
The parameters are
$\Omega_{ab}=\Omega_{bc}=1,Q=0,\omega_{12}=6,\delta=0.3,
\gamma_u=\gamma_v=0.5,\gamma_b=0.15$.
 The two-photon detuning $\Delta$ is plotted in units of $\gamma_u+\gamma_v$.}
        \label{fig:delta}
\end{figure}

The relative magnitude of the central peak to the magnitude of the
side peaks at $\Delta =\pm \omega_{12}/2$ depends on the ratio
$\alpha =\gamma_{b}/\gamma_{a}$. 
In Fig.~\ref{fig:delta} we show the effect of $\alpha$ on the
amplitude of the central peak in the fluorescent intensity on the uv
transition.
 It is seen that the relative amplitude
of the central peak increases with decreasing $\alpha$ (although the 
overall fluorescent intensity decreases). The exact size of the central peak
compared to the side peaks depends on $\Omega$ and $\omega_{12}$ 
as well as $\alpha$. A small value of $\alpha$, which results in a large
central peak as observed in the experiment, is consistent with the 
fact that the decay rates of the intermediate levels are much
smaller than the decay rates of the upper levels~\cite{wang}.
When $\alpha$ increases the central peak becomes relatively smaller, 
and disappears completely for sufficiently large $\alpha$. 
In this case the middle level is 
scarcely populated (because of its large decay rate) and 
the dynamics of the system are dominated by a two-photon process where
the upper levels are directly populated from the ground level.
Thus the large $\alpha$ limit is equivalent to considering only
two-photon processes as in Sec.~III, and it is not surprising that 
the spectrum contains only two peaks as found in that section.

\begin{figure}[tbp]
\center
\centerline{\hbox{
\psfig{figure=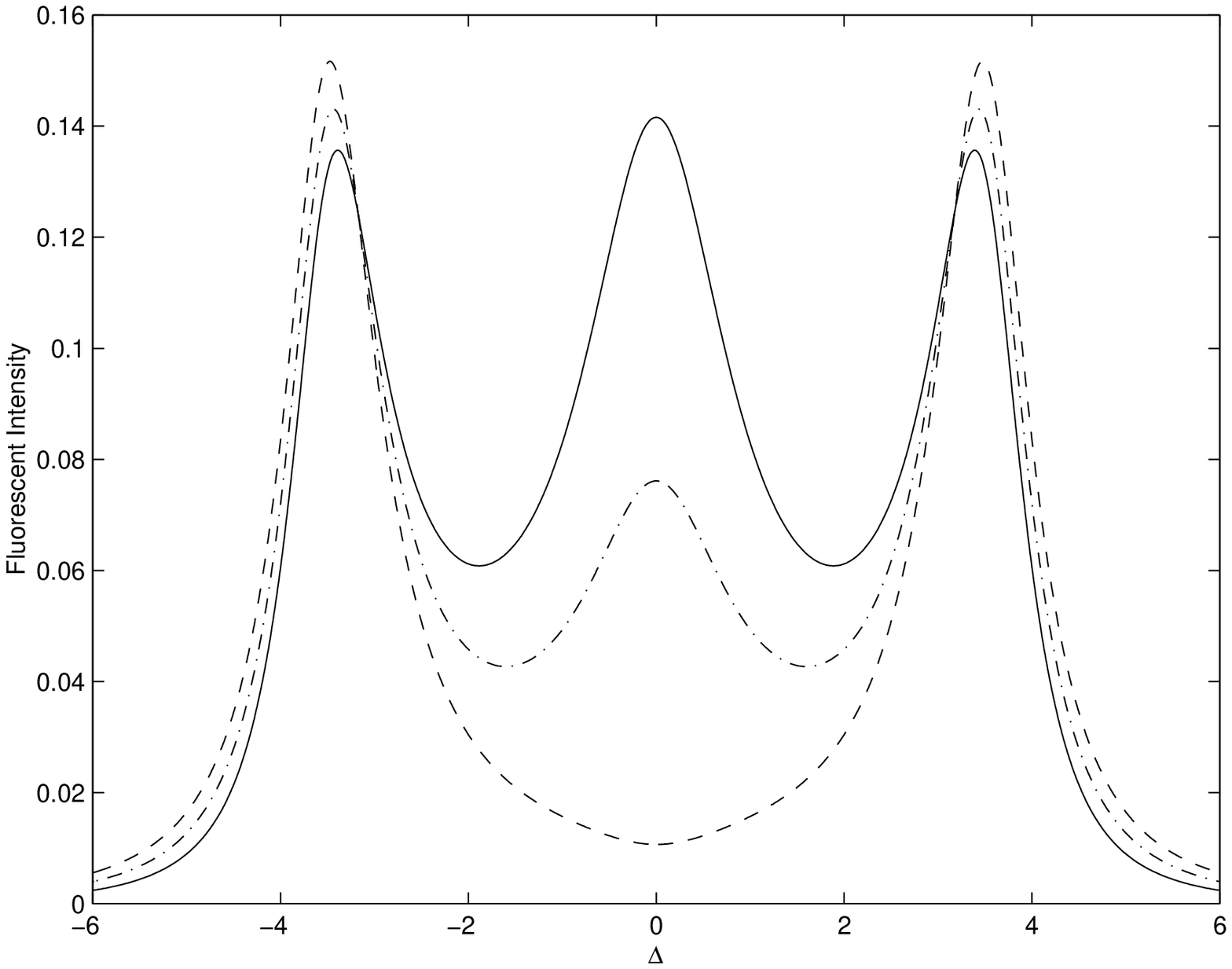,width=5cm}}}     \center

        \caption{Total fluorescent intensity in units of the spontaneous emission
rate
for strong one- and two-photon coupling, with non-equal decay rates on the 
ultraviolet and visibile transitions.
 The different line styles are as in Fig.~\ref{fig:2ph}(b).
The parameters are
$\Omega_{ab}=\Omega_{bc}=1,Q=0,\omega_{12}=6,\delta=0,
\gamma_u=0.7,\gamma_v=0.3,\gamma_b=0.15$.
 The two-photon detuning $\Delta$ is plotted in units of $\gamma_u+\gamma_v
$.}
        \label{fig:uv}
\end{figure}

Finally, in Fig.~\ref{fig:uv} we show numerically that the results are not 
much affected if we relax our previous assumption
that $\gamma_u$ and $\gamma_v$ are equal (with their 
value being denoted by $\gamma_a$). For this plot we choose 
$\gamma_u$ and $\gamma_v$ to be different by more than a factor of two. 
The numerical 
results in this figure, and all of the above figures, 
indicate that the existence of the third peak is a 
robust feature which does not depend upon fine 
tuning of the parameters in the model.

\section{Summary}

We have modeled quantum interference effects in the intensity of the
fluorescence emitted from a five-level molecular system, studied
experimentally by Xia {\it et al.}~\cite{xia}. We have presented an analytical
solution for the fluorescent intensity, valid in the weak-field limit,
and a numerical solution valid for arbitrary strengths of the driving field.
We have been particularly interested in a theoretical explanation of the
experimentally observed dependence of the number of peaks in the fluorescent
intensity on the mutual orientation of the transition dipole moments.
We have assumed that the molecular excitation is composed of a one-step,
two-photon absorption process, and a two-step process involving the
absorption of a single photon in each step. If the excitation is composed
of only the two-photon processes, the fluorescent intensity consists of
two peaks regardless the mutual orientation of the molecular dipole
moments. With the two-step, one-photon processes included, the intensity
consists of two peaks on transitions with parallel dipole moments and
three peaks on transitions with antiparallel dipole moments. This latter case is
in excellent agreement with the experimental observation~\cite{xia}.
The variation of the number of peaks with the mutual polarization of the
dipole moments is a very clear demonstration of quantum interference 
in spontaneous emission.

\section*{Acknowledgments}

This work has been supported by the Australian Research Council, the
University 
of Queensland, Griffith University, and the Department of Employment, Education and Training,
Australia. We
appreciate valuable discussions with G.J. Milburn.

\section*{Appendix}

The complete analytical solution to the upper level coherence in the weak
driving limit with $\gamma_u=\gamma_v=\gamma_a$ is
\bqa
{\rm
Re}\rho_{12}&=&(\Omega_{ab}\Omega_{bc})^2
\left[{\gamma_b\gamma_a(\delta-\Delta/2+\omega_{12}/2)(\Delta
-\omega_{12}/2)+\gamma_b{\gamma_a}^2(\gamma_a+\gamma_b)/{2}}\right.\nn\\
&&+{2{\gamma_a}^2(\delta-\Delta/2+\omega_{12}/2)(\Delta/2+\delta)-\gamma_a({
2\gamma_a+\gamma_b})(\Delta-\omega_{12}/2)(\Delta/2+\delta)}/2\nn\\
&&+\omega_{12}{(\delta-\Delta/2+\omega_{12}/2)(\Delta-\omega_{12}/2)(\Delta/
2+\delta)+\omega_{12}\gamma_a({2\gamma_a+\gamma_b})(\Delta/2+\delta)/2}\nn\\
&&\left.-\omega_{12}\gamma_a{{\gamma_b}(\delta-\Delta/2+\omega_{12}/2)/2+
\omega_{12}{\gamma_b(2\gamma_a+\gamma_b)}(\Delta-\omega_{12}/2)/4}\right]\nn\\
&&\div\left\{[(\delta-\Delta/2+\omega_{12}/2)^2+({2\gamma_a+\gamma_b})^2/4][(
\Delta-\omega_{12}/2)^2+{\gamma_a}^2]\right.\nn\\
&&\left.[(\delta+\Delta/2)^2+{\gamma_b^2}/{4}][4{\gamma_a}^2+{\omega_{12}}^2
]\right\} \nn\\
&&+(\Omega_{ab}\Omega_{bc})^2\left[{{\gamma_b\gamma_a}(\delta-\Delta/2-\omega_{12}/2)(\Delta+\omega_{1
2}/2)/2+{\gamma_b{\gamma_a}^2(2\gamma_a+\gamma_b)}/{2}}\right.\nn\\
&&+{2{\gamma_a}^2(\delta-\Delta/2-\omega_{12}/2)(\Delta/2+\delta)-\gamma_a({
2\gamma_a+\gamma_b}{2})(\Delta+\omega_{12}/2)(\Delta/2+\delta)/2}\nn\\
&&+\omega_{12}{(\delta-\Delta/2-\omega_{12}/2)(\Delta+\omega_{12}/2)(\Delta/
2+\delta)+\omega_{12}({2\gamma_a+\gamma_b})\gamma_a(\Delta/2+\delta)/2}\nn\\
&&\left.-\omega_{12}\gamma_a{{\gamma_b}(\delta-\Delta/2-\omega_{12}/2)/2+
\omega_{12}{\gamma_b(2\gamma_a+\gamma_b)}(\Delta+\omega_{12}/2)/4}\right]\nn\\
&&\div\left\{[(\delta-\Delta/2-\omega_{12}/2)^2+({2\gamma_a+\gamma_b})^2/4][(
\Delta+\omega_{12}/2)^2+{\gamma_a}^2]\right.\nn\\
&&\left.[(\delta+\Delta/2)^2+{\gamma_b^2}/{4}][4{\gamma_a}^2+{\omega_{12}}^2
]\right\}\nn\\
&&+\frac{\Omega_{ab}^2\Omega_{bc}^2[-\gamma_a(2\gamma_a+\gamma_b)-(\delta-
\Delta/2+\omega_{12}/2)\omega_{12}]}{[(\delta-\Delta/2+\omega_{12}/2)^2+
({2\gamma_a+\gamma_b})^2/4][(\delta+\Delta/2)^2+{\gamma_b^2}/{4}]
[4{\gamma_a}^2+{\omega_{12}}^2]}\nn\\
&&+\frac{\Omega_{ab}^2\Omega_{bc}^2[-\gamma_a(2\gamma_a+\gamma_b)-(\delta-
\Delta/2-
\omega_{12}/2)\omega_{12}]}{[(\delta-\Delta/2-\omega_{12}/2)^2+({2\gamma_a+
\gamma_b})^2/4][(\delta+\Delta/2)^2+{\gamma_b^2}/{4}][4{\gamma_a}^2+{\omega_{
12}}^2]}\label{pop12}.
\end{eqnarray}

\end{document}